# In-Context System Identification for Nonlinear Dynamics Using Large Language Models

Linyu Lin

*Abstract*— Sparse Identification of Nonlinear Dynamics (SINDy) is a powerful method for discovering parsimonious governing equations from data, but it often requires expert tuning of candidate libraries. We propose an LLM-aided SINDy pipeline that iteratively refines candidate equations using a large language model (LLM) in the loop through in-context learning. The pipeline begins with a baseline SINDy model fit using an adaptive library and then enters a LLM-guided refinement cycle. At each iteration, the current best equations, error metrics, and domain-specific constraints are summarized in a prompt to the LLM, which suggests new equation structures. These candidate equations are parsed against a defined symbolic form and evaluated on training and test data. The pipeline uses simulation-based error as a primary metric, but also assesses structural similarity to ground truth, including matching functional forms, key terms, couplings, qualitative behavior. An iterative stopping criterion ends refinement early if test error falls below a threshold (NRMSE < 0.1) or if a maximum of 10 iterations is reached. Finally, the best model is selected, and we evaluate this LLM-aided SINDy on 63 dynamical system datasets (ODEBench) and march leuba model for boiling nuclear reactor. The results are compared against classical SINDy and show the LLM-loop consistently improves symbolic recovery with higher equation similarity to ground truth and lower test RMSE than baseline SINDy for cases with complex dynamics. This work demonstrates that an LLM can effectively guide SINDy's search through equation space, integrating data-driven error feedback with domain-inspired symbolic reasoning to discover governing equations that are not only accurate but also structurally interpretable.

## I. INTRODUCTION

Discovering governing equations from time-series data is a central challenge in data-driven modeling of dynamical systems. Sparse regression approaches such as SINDy (Sparse Identification of Nonlinear Dynamics) [[1]] have proven successful at identifying compact analytic laws that explain observed dynamics, by fitting time derivatives to a hand-crafted library of candidate nonlinear features. Classical SINDy requires a priori choosing of a function library and the thresholding of small coefficients to enforce sparsity, both based on human judgement. Such an approach may struggle with complex systems in which the true dynamics involve complex terms outside the basic library.

Recent advances have shown promise by incorporating machine learning and prior knowledge into equation discovery. Beyond the original formulation, a number of SINDy variants have been developed to improve robustness, uncertainty handling, and applicability under limited or noisy data. Weak SINDy replaces pointwise derivative estimation with a weak/Galerkin formulation, substantially improving noise robustness by converting differentiation into integration against test functions[2]. Ensemble-SINDy further improves stability in the low-data/high-noise regime by using bootstrap aggregating to obtain inclusion probabilities over candidate terms, enabling both robust recovery and uncertainty quantification at the level of discovered structures[3]. For control-oriented settings, probabilistic extensions such as Multivariate Gaussian Fit SINDyC (MvG-SINDyC) aggregate ensembles of SINDyC models to quantify uncertainty and improve reliability when simulation-trained models exhibit discrepancies against experimental trajectories [4]. Besides, deep learning models such as neural ordinary differential equations (ODEs) [5] or transformer-based regressors [6], [7] navigate the search space of equations, but often do so as black boxes, without yielding human-readable formulas. Related scientific machine learning approaches also embed physical constraints into neural surrogates (e.g., physics-informed neural networks) to infer latent dynamics and parameters from sparse observations, trading symbolic transparency for function approximation flexibility [8]. Hybrid strategies have also been explored where neural networks expand or learn feature representations that are subsequently sparsified (or otherwise regularized) to regain interpretability, partially mitigating fixed-library limitations [9].

There is growing interest in using large language models (LLMs), which contain broad scientific knowledge, to guide the search for equations. LLMs can suggest relevant functional forms or transformations based on patterns in data or domain context, effectively injecting expert intuition into the search process. At the same time, leveraging domain-specific insights such as data characteristics or known behaviors of function families can prune out implausible candidates and bias the search toward correct solutions.

In this work, we explore augmenting the SINDy algorithm with an LLM to adaptively propose new candidate terms and structures based on data characteristics, model context, and performance, and we clarify how this differs from both SINDy variants and end-to-end learning-based equation discovery. Weak SINDy, Ensemble-SINDy, and probabilistic SINDyC extensions primarily improve robustness and uncertainty quantification given a predefined dictionary class; however, when the governing dynamics lie outside the assumed candidate families, performance can degrade due to library misspecification even if regression and uncertainty estimates are well-behaved [2]-[4]. Our approach targets this complementary failure mode: rather than only stabilizing coefficient recovery within a fixed library, the LLM-in-the-loop proposes new functional families and compositions driven by closed-loop feedback from simulation error and structural constraints. In contrast to deep learning surrogates that can fit trajectories with high expressivity but may not yield compact symbolic laws, we retain a sparse-regression backbone and explicit symbolic representations, so that every refinement remains interpretable and verifiable by simulation. Compared with transformer symbolic regressors that attempt to infer equations directly from trajectories, our pipeline uses the LLM as a constrained hypothesis generator within an explicit validation-and-selection loop, allowing domain constraints and multimodal context (metadata and diagnostic plots) to steer the search while maintaining transparent acceptance criteria. We integrate an LLM into a closed-loop pipeline: after an initial SINDy fit, the LLM receives a summary of the current model's performance and suggests improvements in a structured format. These suggestions are validated and evaluated, and the cycle repeats until a maximum number of iterations is reached. The goal is to leverage the LLM's ability to suggest *novel combinations of features and functional forms* beyond a fixed library, guided by error feedback and domain constraints.

## II. METHOD

We propose an iterative sparse-regression pipeline in which a large language model (LLM) serves as a constrained hypothesis generator that proposes candidate equation structures, while an automated validation-and-selection loop accepts only candidates that (i) can be parsed under a restricted symbolic grammar, (ii) can be fit with linear-in-parameters sparse regression, and (iii) accurately reproduce forward-simulated trajectories on held-out data. The overall procedure alternates between (1) fitting and simulating candidate ODEs, and (2) prompting the LLM with compact diagnostics (data characteristics, baseline behavior, error focus, and prior attempts) to propose locally improved or globally redesigned structures. This mirrors how a domain expert iteratively refines a model using error feedback, but enforces transparent acceptance criteria and repeatable scoring. Fig. 1 shows the schematic of LLM-aided sparse identification workflow. The method is designed to make the search over functional forms explicit, interpretable, and constrained, rather than relying only on sparse symbolic regression:

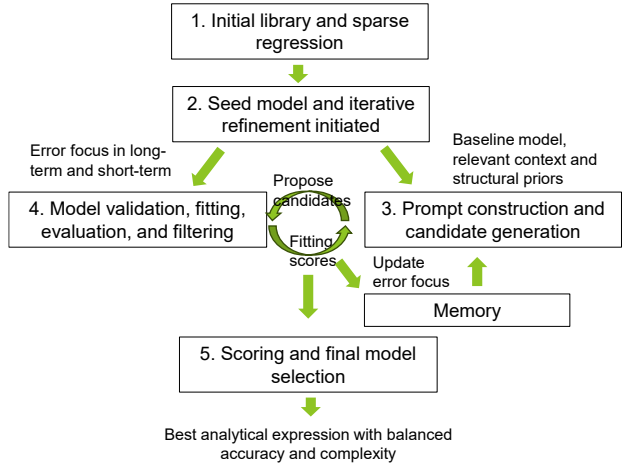

Fig. 1. Schematic of the LLM-aided SINDy workflow

**1. Baseline fit, empirical data characterization, and structural priors:** Given a multivariate trajectory $x(t) \in \mathbb{R}^d$, we first compute compact empirical summaries of each state on the training segment (e.g., range, standard deviation, basic monotonicity/oscillation indicators, approximate period when applicable, and simple saturation heuristics). These statistics are used for two purposes: (i) to populate the LLM prompt with interpretable behavioral cues like "oscillatory" or "saturating", and (ii) to derive lightweight structural priors that guide which function families are preferred, optional, or discouraged for each state derivative. In parallel, we fit an initial "baseline" sparse-regression model using a fixed, broad candidate dictionary to obtain: (i) baseline equations, (ii) train/test derivative-fit scores, (iii) trajectory rollout error on held-out data, and (iv) a trust label per derivative that reflects whether the baseline is numerically reliable for that equation. For example, if the baseline provides a strong fit and stable rollout predictions. These baseline diagnostics are not only used for comparison, they also condition the subsequent refinement loop by informing which families should be emphasized or de-emphasized, and by enabling safeguards during final model selection.

**2. Seed model and iterative refinement protocol:** We initialize the refinement loop with a parsimonious seed structure (typically a minimal set of linear terms), fit its coefficients via sparse regression, and evaluate it using forward simulation on the test interval. The refinement loop then proceeds for up to a fixed iteration budget (e.g., 10) or until an early-stopping criterion is met. At each iteration, we identify an error focus that prioritizes the state(s) with the largest normalized test error, encouraging targeted improvements without destabilizing already well-fit derivatives. We also track short-term improvement; if recent iterations show a plateau with insufficient relative error reduction over a small window, the loop increases exploration by raising LLM sampling diversity and requesting more candidates. This implements a pragmatic exploration–exploitation strategy without changing the underlying sparse-regression backbone.

**3. Prompt construction and constrained candidate generation by the LLM:** At each iteration, the algorithm assembles a structured prompt that encodes the current identification context in a compact, machine-consumable form. The prompt includes the variable and constant metadata, including their names, units, and any known parameter values, empirically derived data-characterization summaries for each state, including range and qualitative behaviors such as oscillation or saturation, and the current-best and baseline equations accompanied by their diagnostic reliability indicators. It also embeds structural priors that express soft preferences over admissible function families for each state derivative, together with a short memory of recently rejected candidates to discourage repetition. Finally, the prompt enforces hard output constraints: the returned candidates must be sparse, must respect the allowed symbolic grammar, and must remain linear in unknown coefficients so that each derivative can be written in the generic form.

$$\dot{x}_i(t) = \sum_{k=1}^{K} \theta_{i,k} \phi_k(x(t), u(t), t) \quad (1)$$

where $\phi_k(\cdot)$ are admissible features that are selected from a constrained library and $\theta_{i,k}$ are scalar coefficients to be fit by sparse regression. The LLM is then asked to produce multiple candidate equation templates in a strictly parseable format so they can be automatically validated, fit, and scored. When available, the prompt may additionally include a compact diagnostic visualization or summary of recent rollout behavior so the LLM can target qualitative failure modes such as phase lag, drift, damping mismatch, or frequency error.

**4. Candidate validation, fitting, simulation-based evaluation, and novelty filtering:** Each candidate produced by the LLM is passed through an automated validation pipeline prior to scoring. The first gate enforces syntactic correctness and grammar compliance, rejecting any proposal that uses disallowed operators or variables, violates the permissible functional forms, or breaks linearity in coefficients. For example, coefficients are not allowed to be put inside nonlinear functions, denominators, or exponents. For candidates that pass syntax, the pipeline enforces sparsity and redundancy constraints by limiting the number of retained terms per equation and discarding candidates, whose term signatures match previously tested structures. Such constraints prevent the pipeline from cycling and improve exploration efficiency. Valid and novel candidates then undergo coefficient estimation on the training segment via sparse regression under the proposed feature structure, after which they are evaluated using forward simulation on the test interval with robust numerical integration settings and a strict wall-clock timeout to prevent pathological models from stalling the run. Candidates that fail to simulate due to divergence, numerical blow-up, or timeout are assigned a dominating penalty with effectively infinite rollout error, ensuring they are not selected in future iterations. For candidates that successfully simulate, the primary metric is the normalized trajectory error computed per state and aggregated via the maximum across states to reflect worst-case predictive fidelity; auxiliary diagnostics such as derivative-level fit can be retained for analysis but do not supersede rollout accuracy in selection.

**5. Multi-objective scoring, early stopping, and final model selection:** Among successfully evaluated candidates in a given iteration, selection is performed using a multi-objective score that balances predictive accuracy with parsimony and prior consistency. Concretely, each candidate is assigned a scalar objective of the form

$$J = \max_i NRMSE_i + \lambda_c \mathcal{C} + \lambda_p \mathcal{P} \quad (2)$$

where $\max_i NRMSE_i$ is the maximum test normalized root mean squared error (NRMSE) over states, $\mathcal{C}$ is a normalized symbolic complexity measure that counts expression-tree node, and $\mathcal{P}$ is a penalty that increases when the candidate violates or unnecessarily proliferates function families relative to the structural priors. This formulation biases the search toward compact, interpretable equations when multiple candidates achieve comparable rollout fidelity. The loop terminates early when the selected candidate achieves a sufficiently low maximum test NRMSE, implemented as a threshold test $\max_i NRMSE_i < \tau$ with $\tau$ set, for example, to 0.1, and otherwise continues until the iteration budget is exhausted. After termination, the algorithm returns the best candidate observed across all iterations; additionally, when the baseline model is labeled reliable, a safeguard is applied to prevent selecting an LLM-refined model that is materially worse than the baseline on held-out rollout error, thereby ensuring the closed-loop process does not regress below a strong initial solution.

Through these steps, our method is intended to balance accuracy and interpretability. The LLM proposals inject flexibility to capture complex behaviors that a fixed library would miss in classical SINDy, while the domain constraints and validation steps guard against nonsense solutions and overfitting. We also incorporate domain-specific knowledge automatically via structural priors, and use simulation error as a feedback channel. Model selection is not purely based on sparsity and error, but also on structural alignment with expectations (via penalties). In essence, classical SINDy yields a single model, which may or may not be correct, whereas our LLM-aided SINDy generates a sequence of models, enabling course-correction and guided exploration of the search space of equations.

## III. EXPERIMENTAL SETUP

We evaluated our approach on ODEBench benchmark [7] and one nuclear-specific problem [11]. The ODEBench consists of 63 diverse ODE systems collected from the literature—specifically, 23 1-D systems, 28 2-D systems, 10 3-D systems, and 2 4-D systems—thus covering a wide range of dynamics. Among these are four known chaotic systems (e.g., variants of Lorenz or other chaotic attractors) as well as many standard nonlinear systems, including predator-prey models, oscillators, and reaction kinetics. Each task provides a reference "ground truth" ODE, which we treated as unknown during our discovery process, along with a set of trajectory data generated from that ODE. In addition, we introduce a March–Leuba boiling water reactor (BWR) reduced-order benchmark [11] that captures coupled neutron kinetics and reactivity feedback effects and is widely used as a surrogate for BWR stability studies.

For each system, we used a single trajectory for training and testing. Typically, we simulated the system over a certain time span and with a given initial condition in order to obtain the state trajectory. We split the data: the first portion of the time series was used to train the model (fitting the ODE), and a later portion (or a separate trajectory from a different initial condition, if provided) served as a test set to evaluate how well the discovered model generalizes and predicts future behavior. For the March–Leuba BWR benchmark, we generated ODEBench-compatible training and test trajectories via closed-loop simulation around distinct setpoint schedules, and exported the control action as an additional exogenous "state-like" regressor to support controlled-dynamics identification while preventing the discovery process from inventing spurious control dynamics. To avoid knowledge leakage, no information other than the numerical simulation data was fed to our algorithms.

We compared our LLM-augmented approach against a baseline symbolic regression that uses a fixed library of candidate terms. No noises are imposed on simulation data. The baseline was configured with a reasonably comprehensive library, including polynomials in each state variable up to a certain degree and common nonlinear functions such as $sin$, $cos$, exponential and logarithm. This baseline employed the same sparse regression algorithm without any library adaptation, which is essentially akin to a standard SINDy approach with an expert-chosen library. Both our method and the baseline shared the same initial library, while LLM continues expanding the function library based on context and model performance, which highlights the value of adaptive term discovery. For the March–Leuba benchmark, we retained the same evaluation protocol but emphasized polynomial and bilinear structure during adaptive refinement to reflect the model's canonical feedback couplings and lag dynamics.

We evaluated performance by using both quantitative error metrics and qualitative structural metrics. For numerical accuracy, we report the coefficient of determination ($R^2$) and NRMSE for the model predictions vs. the ground truth, on both the training data and the test data. A high test $R^2$ (close to 1.0) and a low NRMSE indicate that the model predicts the system's behavior accurately.

For structural recovery, we categorize results as "good" if the derived model captures major terms or behaviors in ground truth equations. This is the most common outcome, with the model able to reproduce the logistic growth's $x$ and $x^2$ terms but adding an extra small term. In other words, the model reflects the right structure but slightly different coefficients. For a good fit, we require at least the correct qualitative behavior or some key term overlap. The structural recovery is "failed" if the model is structurally wrong such that it likely represents a different mechanism. These grades allow us to quantify how often our method not only makes good predictions but also "discovers" the physics instead of just producing a black-box fit. In this manner, we graded both the baseline and our LLM-based method on each of the 63 tasks.

## IV. RESULTS

### A. ODE Benchmark

Across the ODEBench suite, our LLM-augmented approach demonstrated improved prediction accuracy on a majority of the tasks, unlike the fixed-library baseline. Specifically, out of 63 systems, our method achieved early stop criteria with NRMSE = <0.1 on 48 of them, as opposed to 30 for the baseline. The median error for the baseline is around 62% of variable range, whereas for the LLM-aided method it is 3.4%. The error spread is also much tighter with LLM augmentation. Baseline SINDy has five cases with infinite error, but our pipeline managed to find a stable model in all cases. Excluding divergent cases, the mean baseline NRMSE was still 43%, as compared to 3.4% for the LLM-aided method. Figure 2 shows the distribution of the test NRMSEs for the baseline SINDy (orange) and LLM-aided SINDy (blue).

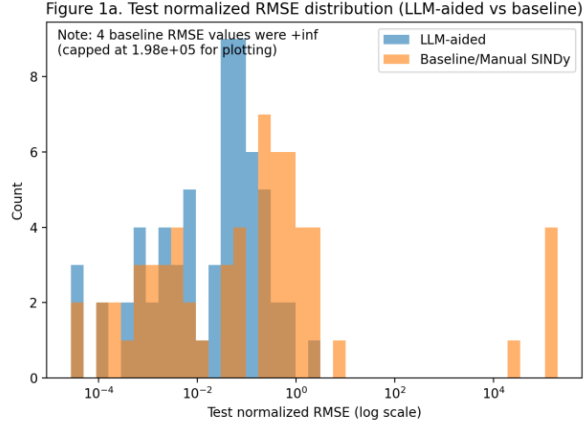

Fig. 2. Test NRMSE distributions comparing the baseline against the LLM-aided SINDy (log-scale). The LLM-aided pipeline achieves substantially lower errors in most cases, as indicated by the shift of the blue distribution toward the lower end.

In terms of structural recovery, the results highlight an important trade-off. The baseline, while often unable to fit the data when its library was insufficient, did recover dominant forms in 26 of the 63 cases. But our LLM-based method managed to recover 36—a substantially higher rate of good structural recovery. The results show that the LLM's guidance increases the chance of finding a model with good predictive accuracy and reasonable complexity, at the expense of sometimes introducing extra terms or different functional forms that deviate from the true physics. The baseline is more binary: it either has the right ingredients for obtaining an exact solution, or it fails and produces nothing useful, since it will not consider functions outside its preset library.

To illustrate the above outcomes, we now discuss a few representative systems.

*Linear oscillator, odebench24:* The ground truth is a 2D mass-spring system without damping $\dot{x} = y, \dot{y} = 2.1x$. Because the data are oscillatory, the baseline included sinusoids and returned a very complicated model: $\dot{x} = 150.93y - 19.78y^3 - 160.36 \sin(y) + 5.21 \sin(2y) + \cdots$ and a similarly lengthy $\dot{y}$. Essentially, the baseline uses many Fourier and polynomial terms to fit the periodic motion. It achieved $R^2 \sim 1$ in training, but the model is not physically correct. On the other hand, the LLM identified this as a harmonic oscillator and proposed linear coupling terms. By iteration 2 it had obtained: $\dot{x} = 0.998y, \dot{y} = -2.098x$, which are exact forms, just with rounding differences. Figure 3 compares the manual and LLM-aided SINDy predictions. This case shows the LLM excelling at interpretation by replacing a baseline fit with the concise true law. Even though the baseline had near-zero error, our selection favored the LLM's simpler model. This hints at the advantage of combining error and complexity in selection—the pipeline did not just settle for the overfit baseline.

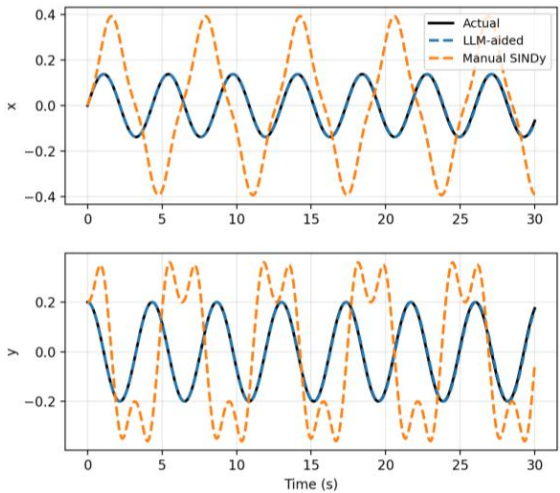

Fig. 3. Manual vs LLM-aided multi-step rollout trajectories

*Exponential forcing, odebench21:* The ground truth is a 1D model with exponential component: $\dot{x} = 1.2 - 0.2x - e^{-x}$. The manual baseline introduces a trigonometric surrogate $\dot{x} = 0.626 x - 0.122 x^2 + 0.230 \cos(x)$ instead of the exponential decay mechanism, leading to a model with wrong structural format but good test metrics test $R^2 \approx 0.975$; NRMSE $\approx 4.9 \times 10^{-3}$. The LLM-aided pipeline starts from a baseline with only polynomial cubic. The LLM's selected candidate explicitly introduced an exponential feature $\dot{x} = -0.8963 x + 0.0634x^2 - 3.1555 \exp(-0.5x) + 3.3728$, aligning with the ground truth's exponential decay. Figure 4 compares the manual and LLM-aided SINDy predictions. The resulting fitted model includes an exponential term and produces stronger test rollouts test $R^2 \approx 0.995$; NRMSE $\approx 2.5 \times 10^{-3}$. However, the recovered exponential is $\exp(-0.5x)$ rather than $\exp(-x)$ and coefficients differ partly due to feature-grid constraints and regression trade-offs, but this case demonstrates a key advantage of LLM guidance: it can pivot the library toward the qualitatively correct mechanism rather than relying on periodic surrogates.

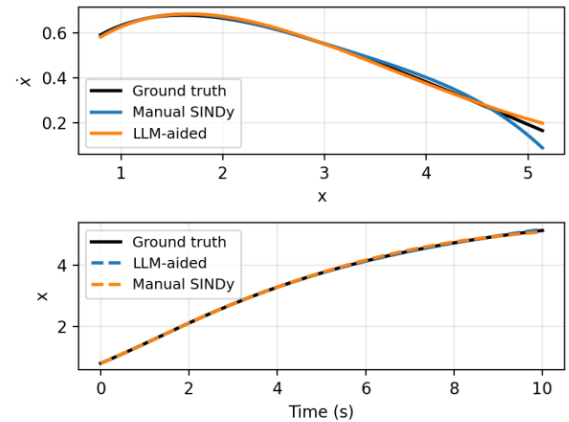

Fig. 4. Manual vs. LLM-aided in predicting $\dot{x}$ vs rollout trajectories.

*SIR-type bilinear coupling, odebench31:* The ground truth features a rational saturation term plus a linear damping: $\dot{x} = -0.4xy$; $\dot{y} = 0.4xy - 0.314y$. The manual model includes incorrect higher-order/trigonometric structure, and the test rollout is unstable with test $R^2 \approx -23.34$ and NRMSE $\approx 0.063/0.082$. From the baseline fitting, LLM is able to realize and suggests that binlinear coupling term of $xy$ is the central driver of $\dot{x}$. LLM also realizes that Fourier terms in $\dot{y}$ are irrelevant as no oscillation behaviors are observed from data characteristics. As a result, the LLM's selected candidate emphasizes the bilinear term and linear decay, yielding a compact model where the learned coefficients closely match ground truth. The test rollout is accurate and stable with test $R^2 \approx 0.994$ and NRMSE $\approx 3.8 \times 10^{-4}$. Here, LLM guidance functions as a structural denoiser by removing weakly supported nuisance terms that the baseline regression included, while preserving the essential coupling.

*Cross-coupled polynomial, odebench39*: The ground truth contains an $x^2y$ term that couples two equations: $\dot{x} = -x + 2.4y +$

$x^2y, \dot{y} = -0.07 - 2.4c + x^2y$. The baseline's polynomial library did include $x^2y$ terms; however, it struggled due to multi-collinearity and ended up with $\dot{x} = -0.892x + 319.36y + 0.968 x^2 y - 204.087x^3 - 0.1\sin(x) + ...$, with $\dot{y}$ being a similar combination of high-order polynomial and Fourier terms. The baseline's simulation results in a very large test $R^2$, indicating poor generalizations. The LLM added the coupling term, and the final model was $\dot{x} = -0.972x + 2.427y, \dot{y} = -2.416x - 0.0342y + 0.070$. Although it still missed the $x^2y$ term, it avoided all spurious trigonometric, exponential, and high-order terms seen in the baseline fit. Metrics improved drastically over the baseline, with test $R^2 = $ ~0.999 vs. the baseline's negative $R^2$, and NRMSE = ~0.007 vs. the baseline's 0.06.

*Rotational shear-flow dynamics, odebench35:* The ground truth is a 2D non-polynomial trigonometric form involving cotangent and squared trigonometric terms: $\dot{x} = \cot y \cos x$; $\dot{y} = \sin x \cos^2 y + 4.2 \sin^2 y$. Both methods exhibit severe test instability with negative test R² and large NRMSE; the LLM-aided run terminates by because maximum number of iterations is reached with test R² ≈ −1.28×10⁵; NRMSE ≈ 1.84, while manual SINDy is even more unstable withtest R² extremely negative; NRMSE ≈ 2.71/2.00. Although it is suggested in the prompt that Fourier/polynomial surrogates should be used. The key limitation is $\cot y$ and products such as $\sin x \cos^2 y$ require either explicit cot and producted-trig features or more expressive compositional bases than the fixed candidate grids provide. The LLM can suggest sin/cos terms, but without the right compositional primitives the regression compensates with many terms and extreme coefficients, which destabilizes rollout.

### B. March-Leuba Reactor Model

The March–Leuba benchmark is a 7-state closed-loop reactor stability model with known lag chains and reactivity coupling, with one channel corresponding to a control action. The LLM prompt includes explicit equation-specific guidance, sensor/actuator lags, and bilinear reactivity coupling in $\dot{x}$. The manual baseline produces an extremely large polynomial model that diverges under rollout, leading to infinite NRMSE and a wildly negative test R². This is consistent with over-parameterized fits in a stiff, feedback-driven setting: small coefficient errors accumulate and destabilize simulation, especially when the control channel is treated as an ordinary state without a consistent exogenous rollout treatment. The LLM-aided run produces a sparse bilinear/polynomial structure for $\dot{x}$ term and linear couplings, with good-looking one-step/fit statistics but poor long-horizon generalization (test R² ≈ 0.871; max NRMSE ≈ 6.68). Figure 5 compares the llm-aided SINDy rollout predictions against ground truth.

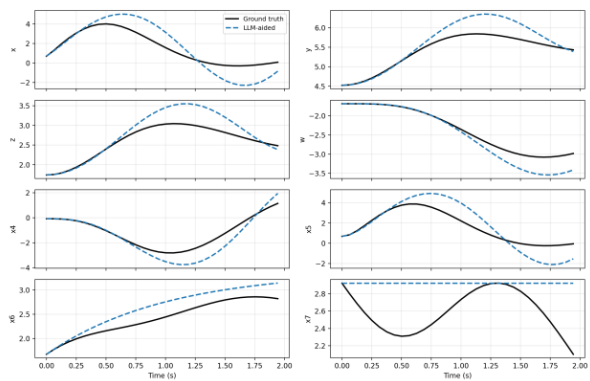

Fig. 5. 2 s rollout trajectories for LLM-aided vs ground truth. Equations from manual diverges at the first simulation step.

Overall, the LLM-aided pipeline significantly outperforms manual SINDy at recovering governing equations in most cases, and also incorporates safeguards to avoid worse solutions. The integration of simulation testing and an LLM-based expert system for proposing new terms is a promising direction for system identification, effectively emulating a human iteration on a model.

### V. CONCLUSION

We introduced an LLM-aided SINDy pipeline that combines sparse regression with LLM-driven symbolic reasoning for nonlinear system identification. Through iterative prompting, the LLM proposes new equation structures that, when validated and fit, lead to more accurate and more interpretable models than afforded by a one-shot SINDy approach. On the 63 benchmark systems, our pipeline improved the rate of good equation recovery and drastically reduced the number of catastrophic failures. It adaptively built libraries containing appropriate functional forms (e.g., polynomial, trigonometric, rational), guided by both data characteristics and LLM insight rather than relying on a fixed library.

This approach effectively automates the work a modeling expert might do: observe model errors, hypothesize new terms, test said terms, and iterate—all within a single algorithmic loop. The LLM serves as a creative generator of hypotheses, while error-driven selection and domain filters ensure that only viable hypotheses are retained. The result is a system identification method that is data-efficient and yields models that align more closely with the underlying physics.

However, the current pipeline is not guaranteed to find the true model in every case. Its success depends on the LLM's ability to suggest relevant terms, as well as on the data being informative enough. In extremely noisy or chaotic systems, it may still fall short. The LLM's suggestions can sometimes overshoot with needless complexity or wrong functional forms, making our validation and selection steps crucial. There is also an implicit reliance on the LLM's training knowledge, with it potentially favoring certain functional forms it "knows" and thereby possibly biasing the search. However, our results show that bias often pushes toward the correct simplicity.

In future works, we will extend this framework to partial differential equations or delay differential equations, with LLMs suggesting spatial derivatives or delay terms. Another avenue is to use multiple LLMs to cross-verify proposals, or to employ a "critic" model to evaluate the physical plausibility of equations. One could also integrate this with laboratory experiment loops, with the LLM perhaps suggesting new experiments to pin down ambiguous terms.

REFERENCES

**REFERENCES**
[1] S. L. BRUNTON, J. L. PROCTOR, and J. N. KUTZ, "Discovering governing equations from data by sparse identification of nonlinear dynamical systems," *Proceedings of the National Academy of Sciences*, **113**(15), pp. 3932-3937 (2016). https://doi.org/10.1073/pnas.1517384113.
[2] D. A. Messenger and D. M. Bortz, "Weak SINDy: Galerkin-Based Data-Driven Model Selection," *Multiscale Modeling & Simulation*, vol. 19, no. 3, pp. 1474–1497, 2021, doi: 10.1137/20M1343166.
[3] U. Fasel, J. N. Kutz, B. W. Brunton, and S. L. Brunton, "Ensemble-SINDy: Robust sparse model discovery in the low-data, high-noise limit, with active learning and control," *Proc. R. Soc. A*, vol. 478, no. 2260, Art. no. 20210904, 2022, doi: 10.1098/rspa.2021.0904.
[4] Seurin, P. and Lin, L., 2026. Uncertainty quantification of a physics-informed model based on sparse identification of a Thermal Energy Distribution System. *Annals of Nuclear Energy*, *226*, p.111865.
[5] R. T. Q. Chen, Y. Rubanova, J. Bettencourt, and D. K. Duvenaud, "Neural Ordinary Differential Equations," *Advances in Neural Information Processing Systems*, **31**, (2018).

https://proceedings.neurips.cc/paper_files/paper/2018/file/69386f6bb1dfed68692a24c8686939b9-Paper.pdf.

[6] M. VASTL, J. KULHÁNEK, J. KUBALÍK, E. DERNER, and R. BABUŠKA, "SymFormer: End-to-End Symbolic Regression Using Transformer-Based Architecture," *IEEE Access*, **12**, pp. 37840-37849 (2024). DOI: 10.1109/ACCESS.2024.3374649.

[7] S. D'ASCOLI, S. BECKER, A. MATHIS, P. SCHWALLER, and N. KILBERTUS, "ODEformer: Symbolic Regression of Dynamical Systems with Transformers," *arXiv preprint arXiv:2310.05573* (2023). https://doi.org/10.48550/arXiv.2310.05573.

[8] M. Raissi, P. Perdikaris, and G. E. Karniadakis, "Physics-informed neural networks: A deep learning framework for solving forward and inverse problems involving nonlinear partial differential equations," *J. Comput. Phys.*, vol. 378, pp. 686–707, 2019, doi: 10.1016/j.jcp.2018.10.045.

[9] K. Champion, B. Lusch, J. N. Kutz, and S. L. Brunton, "Data-driven discovery of coordinates and governing equations," *Proc. Natl. Acad. Sci. U.S.A.*, vol. 116, no. 45, pp. 22445–22451, 2019, doi: 10.1073/pnas.1906995116.

[10] P. SHOJAEE, K. MEIDANI, A. B. FARIMANI, and C. K. REDDY, "Transformer-based Planning for Symbolic Regression," *Advances in Neural Information Processing Systems*, **36**, pp. 45907-45919 (2023). https://proceedings.neurips.cc/paper_files/paper/2023/file/8ffb4e3118280a66b192b6f06e0e2596-Paper-Conference.pdf.

[11] March-Leuba, J., 1986. A reduced-order model of boiling water reactor linear dynamics. *Nuclear Technology*, *75*(1), pp.15-22.